\documentclass[12pt,a4paper]{article}

\usepackage[]{graphicx}
\usepackage{amsmath,amssymb}
\usepackage{cite}
\usepackage{subfigure}
\usepackage{color}

\def\BibTeX{{\rm B\kern-.05em{\sc i\kern-.025em b}\kern-.08em
		T\kern-.1667em\lower.7ex\hbox{E}\kern-.125emX}}
\usepackage{booktabs}
\usepackage{multirow}
\usepackage{array}
\newcolumntype{+}{>{\global\let\currentrowstyle\relax}}
\newcolumntype{^}{>{\currentrowstyle}}

\newcommand{\VDS}{\ensuremath{\textrm{V}_{DS}}}
\newcommand{\RDS}{\ensuremath{\textrm{R}_{DS}}}
\newcommand{\Dit}{\ensuremath{\textrm{D}_{it}}}
\newcommand{\VGS}{\ensuremath{\textrm{V}_{GS}}}
\newcommand{\PS}{\ensuremath{\textrm{P}_{S}}}
\newcommand{\PT}{\ensuremath{{P}_{T}}}

\newcommand{\IDS}{\ensuremath{\textrm{I}_{DS}}}

\begin{document}
	\onecolumn
	\setlength{\abovedisplayskip}{3.5pt}
	\setlength{\belowdisplayskip}{3.5pt}
	
	\pagenumbering{gobble}
	
\title{Operation and Design of Ferroelectric FETs \\for a BEOL Compatible Device Implementation}
\author{Daniel Lizzit and David Esseni\\
	\small{DPIA, University of Udine,
	via delle Scienze 206, 33100 Udine, Italy}\\
	\small{e-mail: daniel.lizzit@uniud.it}
}
\date{\vspace{-5ex}}
\maketitle

\begin{abstract}
We present a study based on numerical simulations and comparative analysis of recent experimental data concerning the operation and design of FeFETs. Our results show that a proper consideration of charge trapping in the ferroelectric-dielectric stack is indispensable to reconcile simulations with experiments, and to attain the desired hysteretic behavior of the current-voltage characteristics. Then we analyze a few design options for polysilicon channel FeFETs and, in particular, we study the influence of the channel thickness and doping concentration on the memory window, and on the ratio between the polarization dependent, high and low resistance state.
\end{abstract}
\textbf{© 2021 IEEE. Personal use of this material is permitted. Permission from IEEE must be obtained for all other uses, in any current or future media, including reprinting/republishing this material for advertising or promotional purposes, creating new collective works, for resale or redistribution to servers or lists, or reuse of any copyrighted component of this work in other works.\\
	\\
https://ieeexplore.ieee.org/document/9631764}

\section{Introduction}
\label{Sec:Intro}
%
The slowing down of the CMOS geometrical scaling has steered the electron devices research to new functionalities for novel computational paradigms, as well as to the energy efficiency \cite{Review_Seabaugh,Esseni_SST2017}. In particular, there is a growing interest for memristors capable of multiple resistance levels, that have intriguing applications in crossbar arrays for artificial deep neural networks \cite{Ambrogio_Nature2019}, and in hybrid memristive-CMOS circuits for neuromorphic computing \cite{Yu_IEEE_Proc2018,Chicca_APL2020}.

The discovery of ferroelectricity in hafnium oxides opened new perspectives for ferroelectric CMOS devices \cite{Boscke_IEDM2011}, with applications ranging from negative capacitance transistors \cite{Salahuddin_NL2008,Rollo_Nanoscale2020}, to non volatile memories and memristors \cite{Mikolajick_IEDM2019,Slesazeck_Nanotechnology2019}. In particular, Ferroelectric FETs (FeFETs)
with multiple resistance levels have been already reported in an industrial technology \cite{Mulaosmanovic_EDL2020}. However, in order to unleash the potentials of FeFETs as synaptic devices, a Back-End-Of-Line (BEOL) compatible device architecture is in high demand. In fact the BEOL fabrication of FeFETs right on top of CMOS circuits holds the promise for great advantages in terms of performance and energy dissipation.

Because the polarization switching dynamic in FeFETs and the involved charge components are still debated \cite{Toprasertpong_IEDM2019,Toprasertpong_VLSI2020,Deng_IEDM2020}, the operation and design of BEOL compatible FeFETs is still a stimulating and challenging research topic.

We here present a simulation based study that first investigates the operation of experimentally reported FeFETs \cite{Mulaosmanovic_EDL2020}, and then leverages the physical insight to discuss, on a sound ground, the design of BEOL compatible FeFETs.

\begin{figure}[h!]
	\centering
	\includegraphics[width=0.8\hsize]{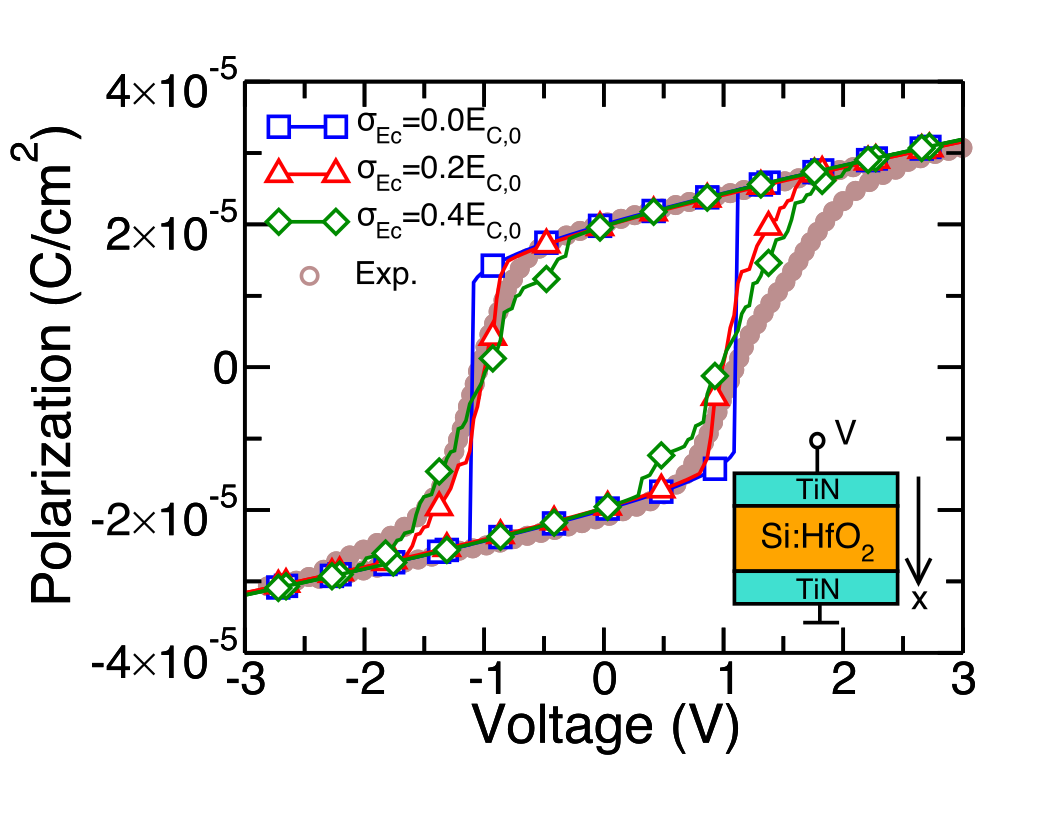}
	\vspace{-4mm} \caption{
		Total polarization \PT\ in a 10 nm Si doped HfO$_2$ MFM capacitor. Simulations were obtained with 100 ferroelectric grains and for different values of $\sigma_{Ec}$. Landau’s coefficients for the coercive field  E$_{C,0}$ (i.e. for $\sigma_{Ec}$=0) are $\alpha$=$-5.37$$\times 10^8$ m/F, $\beta$=$9.62$$\times 10^8$ m$^5$/(FC$^2$), $\gamma$=$9.59$$\times 10^{10}$ m$^9$/(FC$^4$).
		The relative dielectric permittivity of the ferroelectric oxide is $\varepsilon _{r,F}$=30 \cite{Muller_ECS2015}.		
		The tilting of the polarization switching branch depends on the dispersion of the coercive field. Experiments are from \cite{Trentzsch_IEDM2016}.
	}
	\label{Fig:PVcalibration}
\end{figure}
%
\section{Ferroelectric models and calibration}
\label{Sec:TCAD simulator}

Device simulations were carried out using the Sentaurus-Device (S-Device) TCAD tool relying on drift-diffusion transport models and accounting for polarization effects within the Ginzburg–Landau–Khalatnikov framework \cite{SdeviceManual}.

We here assume the total polarization to be separated into the spontaneous polarization \PS\ and a background contribution \cite{Tagantsev_PRB2008, Woo_APA2008},
described by the dielectric permittivity of the ferroelectric material. 
Because thin Hf-based ferroelectrics are tipically in a polycristalline state, and TEM/SEM images evidence grains that have a fairly uniform polarization \cite{Fengler_AEM2017,Park_Nanoscale2017,Mulaosmanovic_AMI2017}, in our simulations we also introduced  ferroelectric grains separated by a 5\AA\ thick oxide, which is consistent with the size of the domain wall region reported in  \cite{Hoffmann_Nanoscale2018}. 

The calibration of Landau's anisotropic constants $\alpha$, $\beta$ and $\gamma$ has been performed by comparing simulations against experimental data for an MFM capacitor. In Fig.\ref{Fig:PVcalibration} we show the total polarization \PT $=$ \PS $+ \varepsilon_0 \varepsilon_{r,F}E_{F}$  in the $x$ direction and obtained under quasi-static triangular voltage ramps from -3 to +3V.
To include the grain to grain variabily for the $\alpha$, $\beta$ and $\gamma$ parameters in such large area devices, we considered a normal distribution of the coercive field around the mean value (E$_{C,0}$) with standard deviation $\sigma_{E_C}$.
The resistivity for the ferroelectric switching employed in simulations is $\rho$$=$30 $\Omega$$\cdot$m \cite{Kobayashi_IEDM2016}, resulting in a time constant $\tau$=$\rho/2\left| \alpha \right|$=28 ns such that the ferroelectric operates in quasi-static conditions for all the bias waveforms explored in this paper. We see in Fig.\ref{Fig:PVcalibration} that a fairly good agreement with experimental results is obtained for $\sigma_{E_C}$=0.4E$_{C,0}$.
 
\section{Operation of experimentally reported FeFETs}
\label{Sec:Analysis of previous experimental results}
\begin{figure}
	\centering
	\includegraphics[width=0.6\hsize]{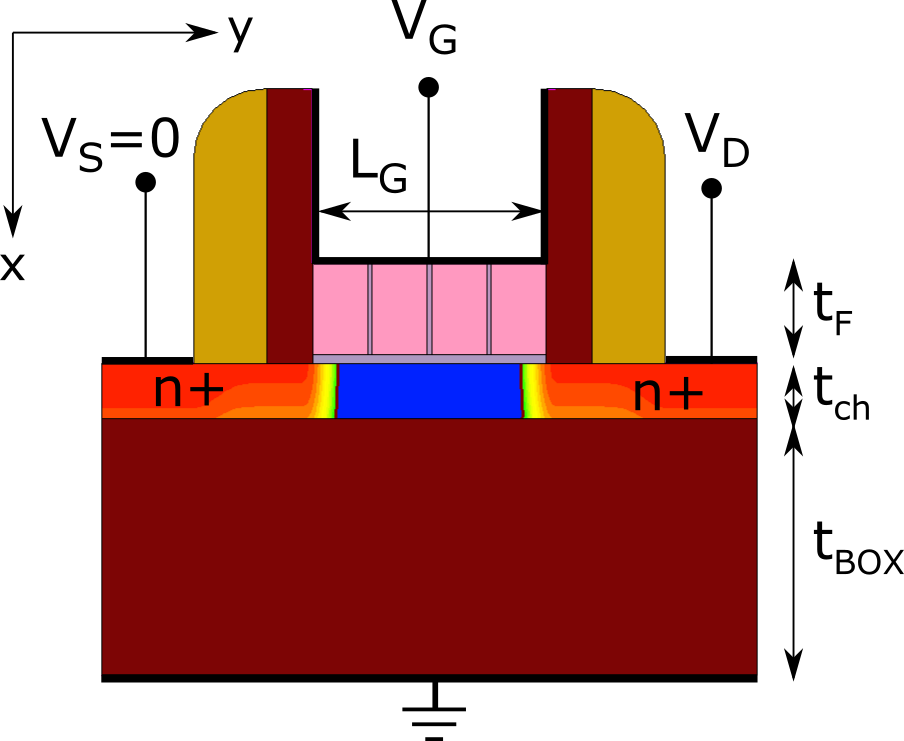}
	\vspace{-3mm}
	\caption{ Sketch of the simulated FeFET. The gate length is $L_G$=24 nm and it is simulated by using 4 grains of 6 nm length. The thickness of the Si:HfO$_2$ and of the interfacial SiO$_2$ layer are $t_{F}$=10nm and $t_{IL}$=1nm, respectively \cite{Dunkel_IEDM2017, Trentzsch_IEDM2016}. The silicon channel has a thickness $t_{ch}$=6 nm, while the SiO$_2$ back oxide is 30nm thick. Each ferroelectric grain of length 6 nm is separated by the other grains by a 5 \AA\ thick non ferroelectric oxide with the same relative permittivity of the Si:HfO$_2$ ferroelectric. The source and drain n-doping are 5$\times$10$^{20}$ cm$^{-3}$ and the Si channel is undoped.
	 }
	\label{Fig:FeFETsketch}
\end{figure}

We here report an analysis of the experimental results recently reported for fully depleted SOI FeFETs based on a silicon doped HfO$_2$ (Si:HfO$_2$) \cite{Dunkel_IEDM2017,Mulaosmanovic_EDL2020,Trentzsch_IEDM2016}; a sketch of the device is shown in Fig.\ref{Fig:FeFETsketch}.
Our main goal is to reproduce the qualitative features of the measured \IDS -\VGS\ curves, so as to gain an understanding of the device operation that will be used in Sec.\ref{Sec:BEOL FeFETs} for the design of BEOL compatible FeFETs.
In order to describe the mobility in the fully depleted SOI transistors we used the thin-layer mobility model \cite{SdeviceManual,Reggiani_TED2007}, that accounts  for the mobility degradation due to the vertical electric field, and also for the scattering due to random fluctuations of the channel thickness and to surface phonons \cite{Book_NanoscaleMOSTransistors,SdeviceManual}.
\begin{figure}
	\centering
	\includegraphics[width=0.80\hsize]{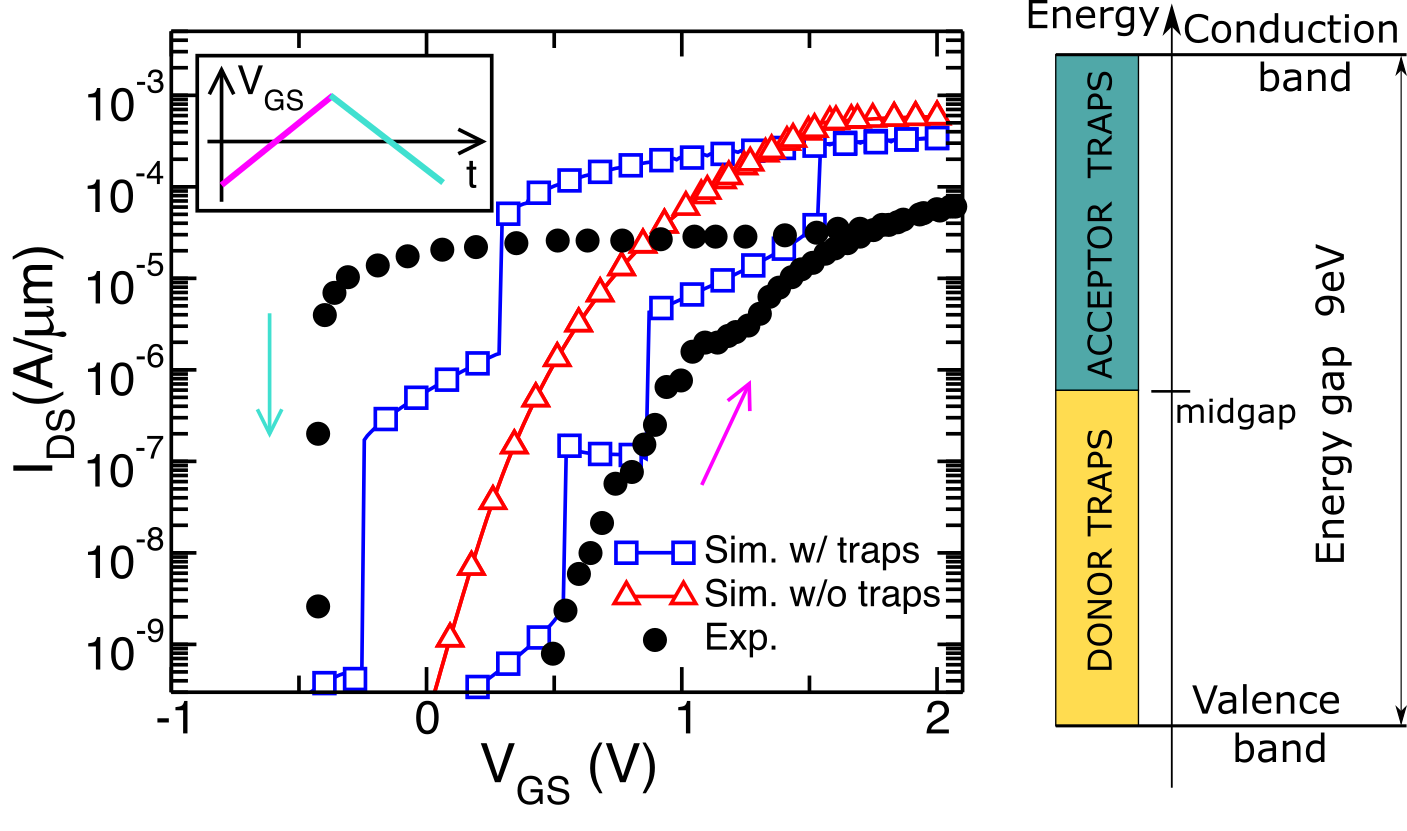}
	\vspace{-2mm} \caption{{\it (left)} \IDS $-$\VGS\ at a \VDS\ of 50mV consistent with the experiments from \cite{Mulaosmanovic_EDL2020}.
		Simulations obtained with and without border traps.	 Experimental results  have been horizontally shifted to ease comparison against simulations.  The inset shows the forward and backward quasi-static gate sweeps used in simulations with a slew rate of $\approx$1V/s and peak-to-peak value of 3V consistently with experiments \cite{Mulaosmanovic_EDL2020}.
		{\it (right)} Acceptor and donor border traps with uniform density energy distribution in the SiO$_2$ IL as used throughout this work. \Dit $_{,acc}$=8$\times 10^{20}$ eV$^{-1}$cm$^{-3}$ and \Dit $_{,don}$=4$\times 10^{20}$ eV$^{-1}$cm$^{-3}$. Traps are also uniformly distributed across the 1nm thick IL. }
	\label{Fig:Ids_Vgs_FDSOIFEFET}
\end{figure}
\begin{figure}
	\centering
	\includegraphics[width=0.99\hsize]{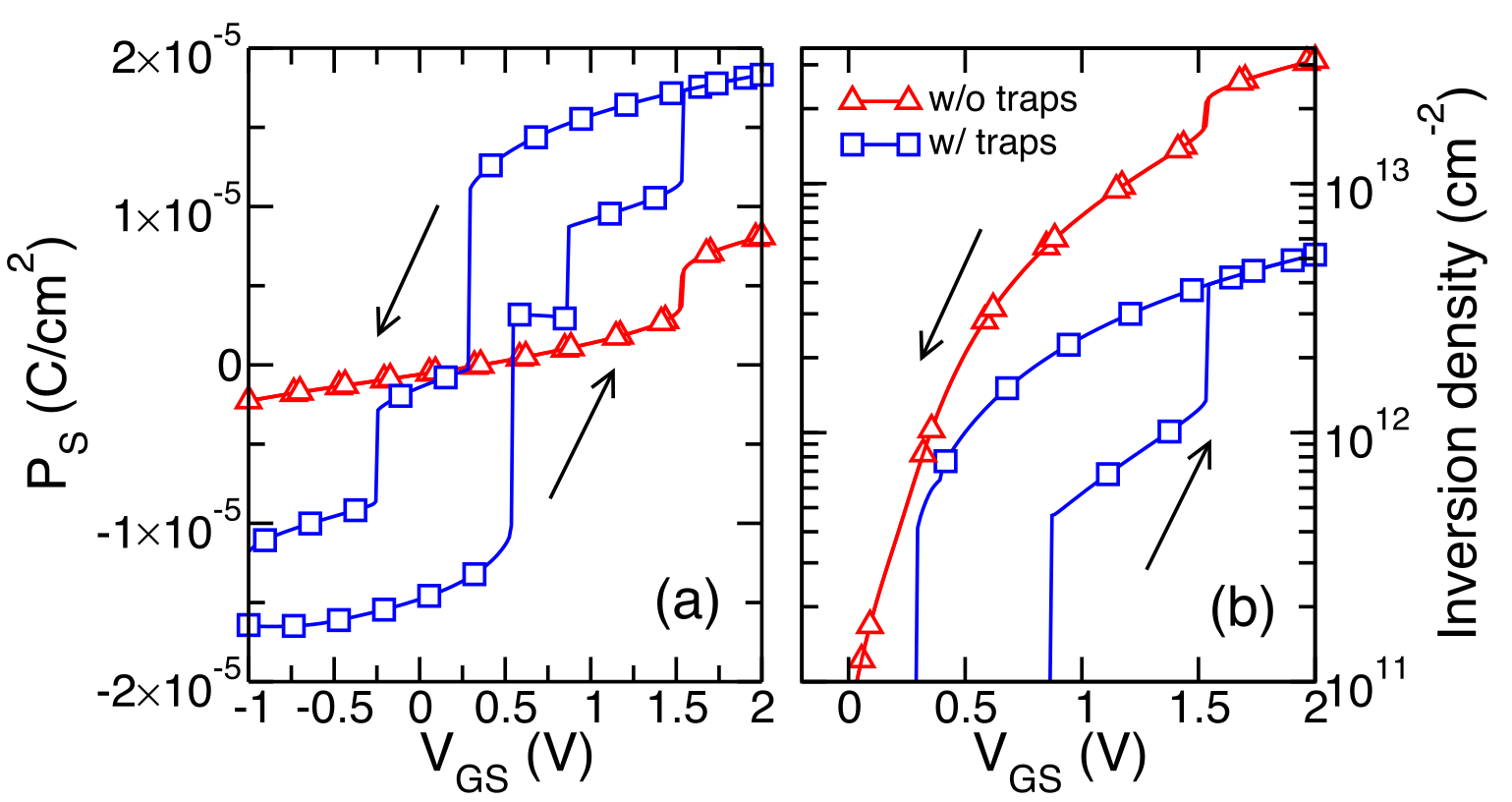}
	\vspace{-2mm} \caption{(a)
		Spontaneous polarization as a function of the gate voltage.
		(b)
		Semiconductor electron density obtained by integrating the volumetric free electrons in the channel and by dividing for the channel length.}
	\label{Fig:PV_withAndwithoutTraps}
\end{figure}
In Fig.\ref{Fig:Ids_Vgs_FDSOIFEFET} we compare the simulated and experimental \IDS\ versus \VGS\ curves \cite{Mulaosmanovic_EDL2020}. The simulations neglecting any trapped charge (either at the ferroelectric-dielectric interface or in the dielectrics) exhibit essentially no hysteresis (red triangles), which is in stark disagreement with experiments; the lack of hysteresis is confirmed in corresponding $P_S$ versus \VGS\ curve in Fig.\ref{Fig:PV_withAndwithoutTraps}(a). We could find the onset of a very limited hysteretic behaviour by enlarging the simulated \VGS\ swing well above the 3V swing used in experiments (not shown), but still the hysteresis is much smaller than in the measured \IDS-\VGS\ curve.

In order to understand and overcome this large discrepancy between simulations and experiments we recall the experimental results in \cite{Toprasertpong_IEDM2019}, which combined quasi-static split C-V measurements with Hall measurements to demonstrate that in bulk FeFETs only a small fraction of the ferroelectric polarization can be compensated by the free-carriers in the transistor channel, whereas most of the charge stems from trapping in the interfacial layer or at the ferroelectric-dielectric interface. In our simulation setup we described the charge trapping by using border traps in the interfacial SiO$_2$ layer (IL) oxide. Such traps are assumed to exchange carriers with the transistor channel via the tunnelling, as described by the Nonlocal Tunneling model accounting for both elastic and inelastic phonon-assisted processes \cite{SdeviceManual}.

Because we feel that there is no unique combination of traps that can be inferred from a comparison to experiments, after several tests we pragmatically opted for a simulation setup consisting of a donor and acceptor type trap density that is spatially and energetically uniform in the interfacial layer (see the sketch in Fig.\ref{Fig:Ids_Vgs_FDSOIFEFET}).

The influence of traps is illustrated in Fig.\ref{Fig:Ids_Vgs_FDSOIFEFET}, where a hysteretic window in the \IDS -\VGS\ curve comparable to experimental results can eventually be obtained (blue squares). In fact the trapped charge in the IL oxide tends to partly compensate the ferroelectric polarization, thus reducing the depolarization and opening a hysteresis loop, that can be also seen in the \PS -\VGS\ plot reported in Fig.\ref{Fig:PV_withAndwithoutTraps}(a).

The inclusion of border traps in the IL helps reconcile simulations with experiments in several respects. In the absence of traps, the ferroelectric polarization results in unphysically large carrier inversion densities (see Fig.\ref{Fig:PV_withAndwithoutTraps}(b)), and in a huge electric field in the IL (not shown).
In the presence of traps, instead, the channel inversion density becomes much smaller than the trapped charge, which is in qualitative agreement with the findings in \cite{Toprasertpong_IEDM2019}.

By comparing Fig.\ref{Fig:PV_withAndwithoutTraps}(b) and \ref{Fig:Ids_Vgs_FDSOIFEFET} one can see that the strong reduction of inversion density caused by the inclusion of traps does not lead to a commensurate \IDS\ reduction. This is because the border traps reduce the transverse electric field in silicon, which improves the mobility and partly compensates for the loss of inversion charge.

We finally notice that the steps in the simulated \IDS\ versus \VGS\ curves in Fig.\ref{Fig:Ids_Vgs_FDSOIFEFET} are due to the switching of a single ferroelectric grain. This effect is most probably exaggerated by the 2D simulation setup (here employed to reduce the computational burden), which neglects the multiple conduction paths possibly occurring in an actual 3D device.


\section{Design of BEOL compatible FeFETs}
\label{Sec:BEOL FeFETs}

\begin{figure}
	\centering
	\includegraphics[width=0.7\hsize]{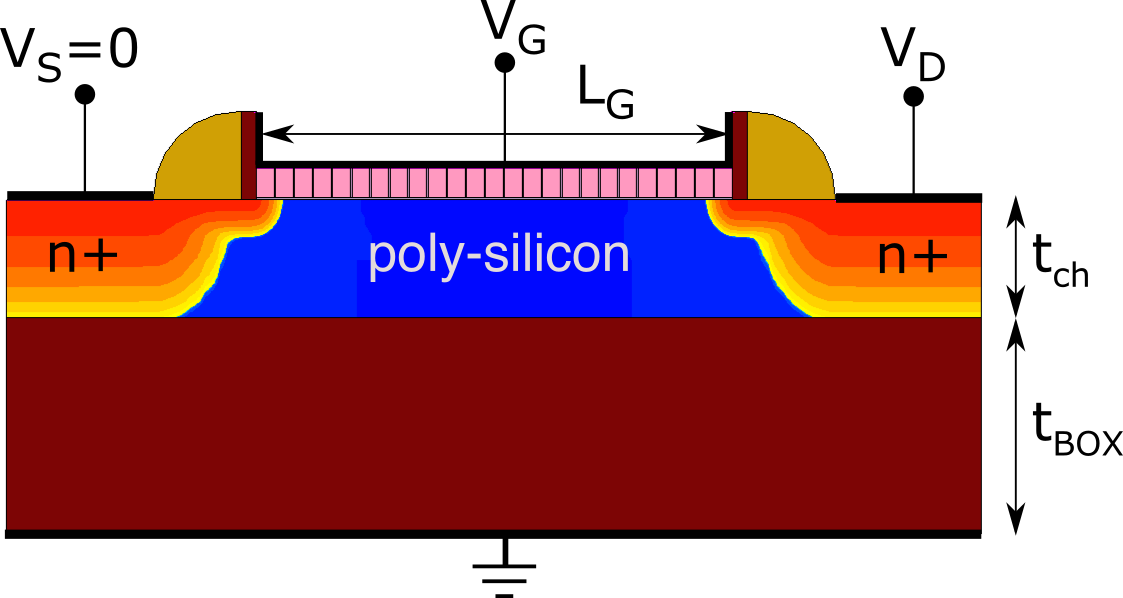}
	\vspace{-2mm}
	\caption{
		Cross sectional view of the simulated FeFET with L$_{G}$=150nm and t$_{ch}$=40nm. The polycrystalline ferroelectric oxide consists of 25 grains with a length of 6 nm and a 5 $\AA$-thick spacer between each grain has been used. Anisotropic constants were obtained with $\sigma_{E_C}$=0.4E$_{C,0}$ consistently with Fig.\ref{Fig:PVcalibration}. The IL SiO$_2$ is 1 nm thick. Source and drain donor doping concentrations are 5$\times$10$^{20}$cm$^{-3}$, whereas the channel acceptor/donor doping ranges between 10$^{16}$ and 10$^{18}$ cm$^{-3}$.
		The silicon back oxide has a thickness t$_{BOX}$=200 nm.
	}
	\label{Fig:BEOL_FeFET}
\end{figure}

\begin{figure}
	\centering
	\includegraphics[width=.9\hsize]{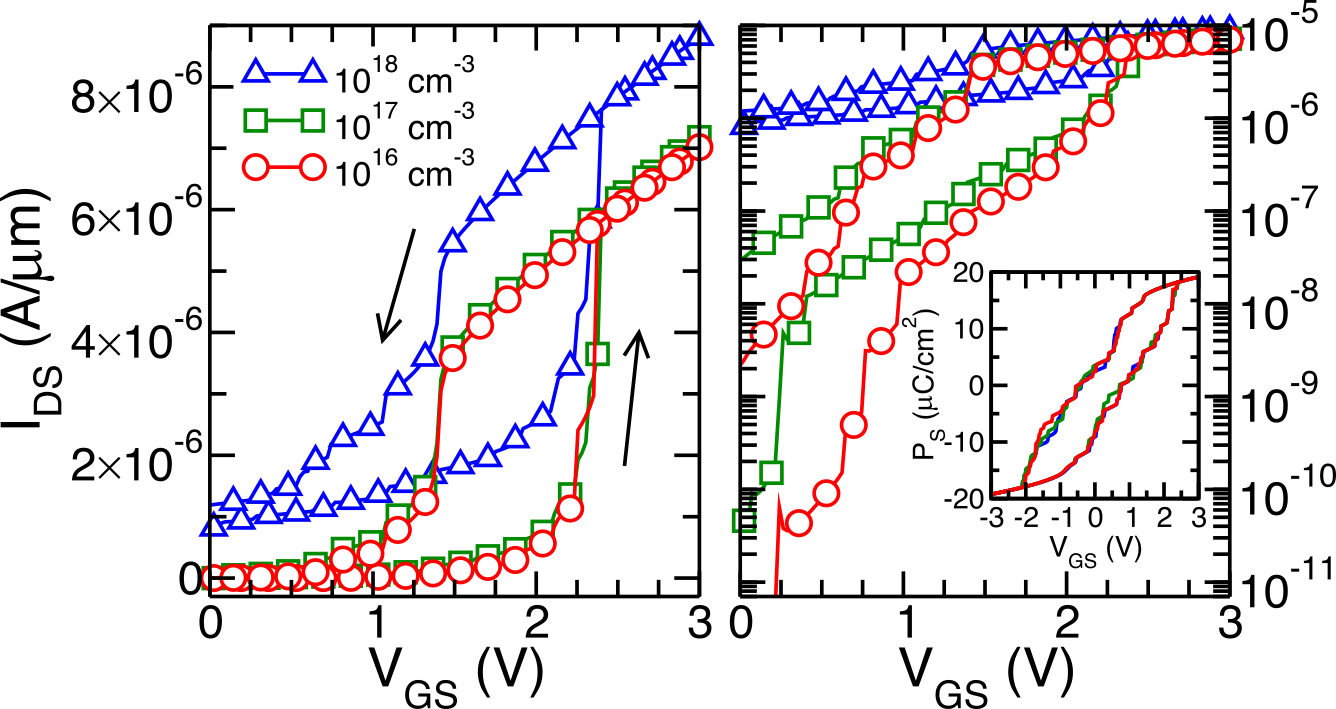}
	\vspace{-7mm}
	\caption{\IDS -\VGS\ curves for a depletion-mode FeFET and for different channel doping concentrations for a  quasi-static gate sweep with \VGS\ from -3 to +3V. The channel thickness is t$_{ch}$=40nm. Results are shown either in linear or in semilogarithmic scales and for \VDS =50mV.}
	\label{Fig:Ids_DepletionModee}
\end{figure}

\begin{figure}
	\centering
	\includegraphics[width=.9\hsize]{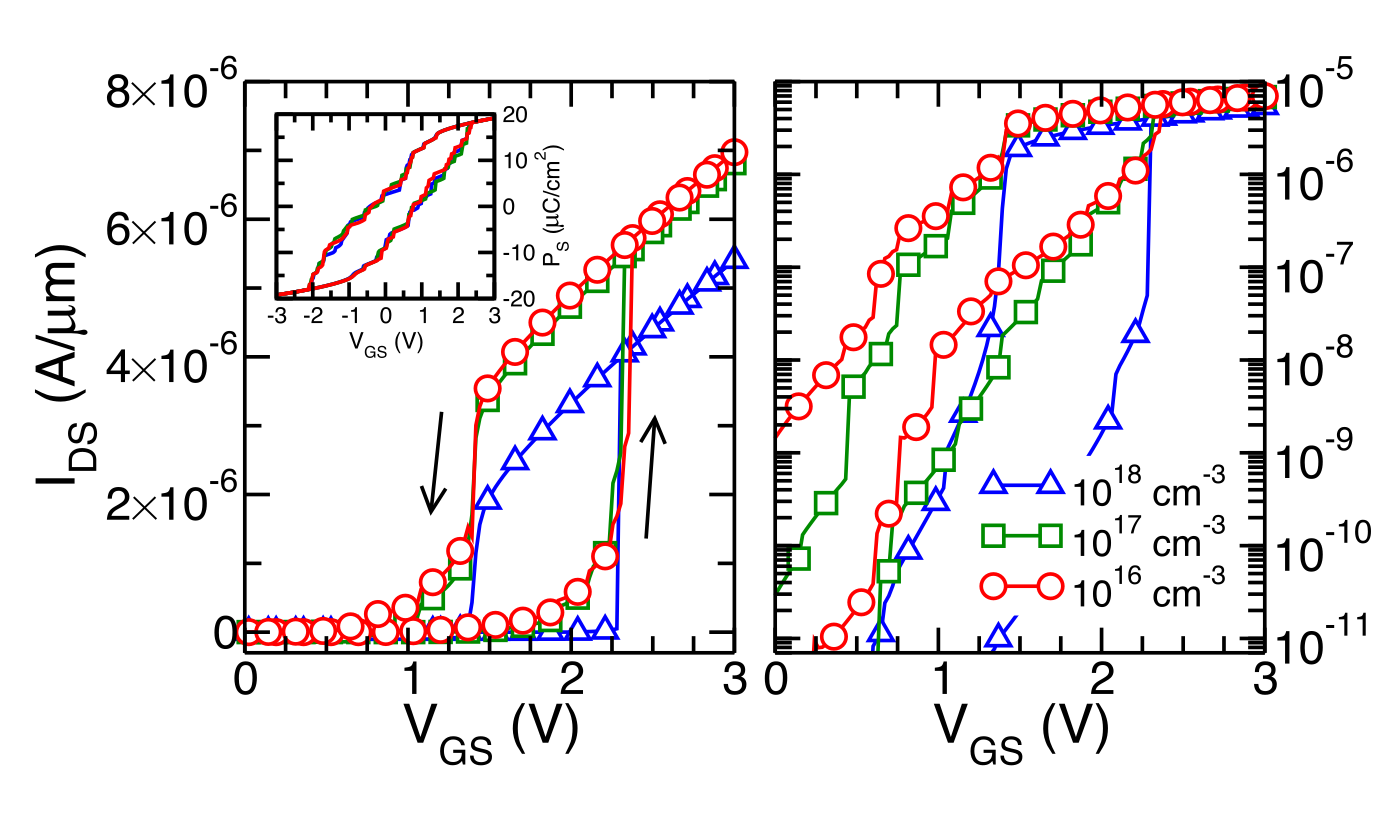}
	\vspace{-7mm}
	\caption{\IDS -\VGS\ curves for an enhancement-mode FeFET and for different channel doping concentrations for a  quasi-static gate sweep with \VGS\ from -3 to +3V. The channel thickness is t$_{ch}$=40nm. Results are shown either in linear or in semilogarithmic scales and for \VDS =50mV.}
	\label{Fig:Ids_EnhancementMode}
\end{figure}

In this section we consider different design options for a polysilicon channel FeFET compatible with a BEOL integration, whose device structure is illustrated in Fig.\ref{Fig:BEOL_FeFET}.
Our analysis will be carried out by employing the same trap distributions as for the SOI FeFET in Sec.\ref{Sec:Analysis of previous experimental results}, that ensure a hysteretic \IDS -\VGS\ curve irrespective of the device thickness and channel doping concentration (see also Figs.\ref{Fig:Ids_DepletionModee}, \ref{Fig:Ids_EnhancementMode}).

We studied both a depletion and an enhancement mode version of the device in Fig.\ref{Fig:BEOL_FeFET}, and for both options we considered channel doping concentrations ranging from 10$^{16}$ cm$^{-3}$ up to 10$^{18}$  cm$^{-3}$ and two different channel thicknesses, namely t$_{ch}$=40 and 80 nm. For the polysilicon thin-film channel we assumed a constant carrier mobility of 10 cm$^{2}$/Vs \cite{Lifshitz_TED1994}.
The channel length is $L_G$$=$150 nm and we studied the \IDS\ and resistance \RDS=\VDS$/$\IDS\ at \VDS$=$50mV. The carrier mobility and the longitudinal field (\VDS$/$$L_G$) are such that velocity saturation effects are negligible for the results of this section.
In Fig.\ref{Fig:Ids_DepletionModee} we see that large doping concentrations in the depletion-mode FeFETs tend to increase \IDS\ and, as expected, to reduce the \IDS\ modulation between the forward and backward quasi-static gate sweep.
On the other hand the size of the memory window is approximately insensitive to the doping concentration. This is because the \PS\ versus \VGS\ hysteresis loop is mainly governed by the border traps in the IL, and it is thus fairly independent of the channel doping concentration, as illustrated in the inset of Fig.\ref{Fig:Ids_DepletionModee}.
Fig.\ref{Fig:Ids_EnhancementMode} shows that the \PS\ hysteresis loop is independent of the channel doping concentration even in the enhancement-mode polysilicon FeFETs. In these latter devices, however, a large doping reduces the \IDS\ values and enlarges   the \IDS\ modulation between the forward and backward quasi-static \VGS\ sweep.

\begin{figure}
	\centering
	\includegraphics[width=.85\hsize]{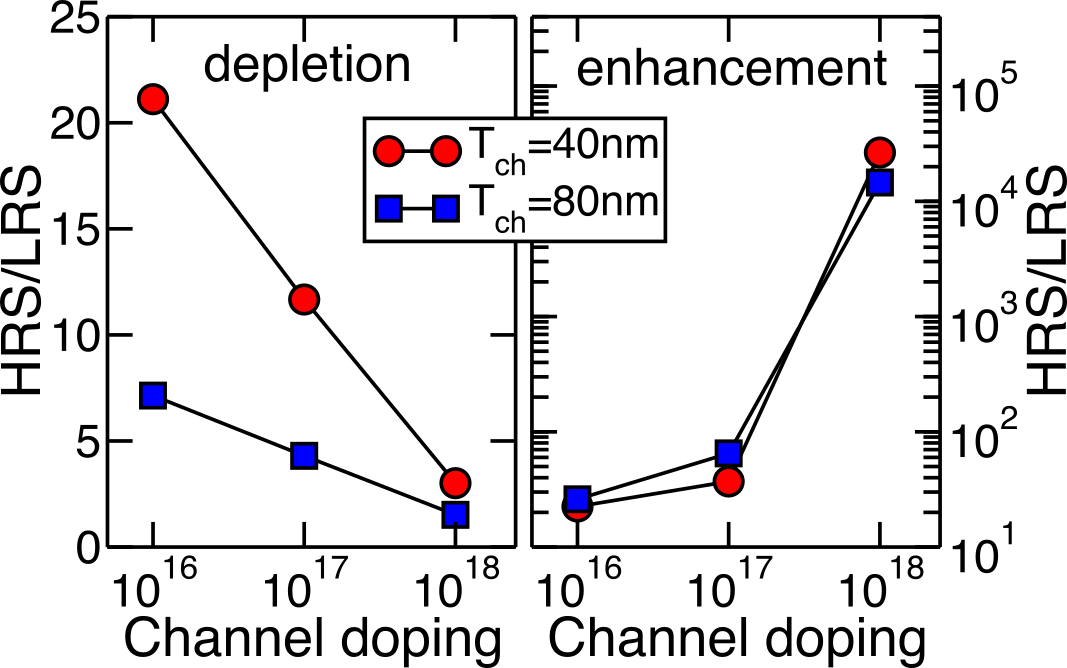}
	\vspace{-4mm} \caption{HRS/LRS ratio as a function of the channel doping and channel thickness and for enhancement- and depletion-mode FeFETs.}
	\label{Fig:ResistanceRatios}
\end{figure}

The simulation results in Figs.\ref{Fig:Ids_DepletionModee} and \ref{Fig:Ids_EnhancementMode} were then used to determine the high resistive state (HRS) and low resistive state (LRS) value, evaluated at \VDS$=$50mV and \VGS\ of 1.75V, namely at the \VGS\ value approximately corresponding to the center of the \IDS\ memory window.
The ratio between the HRS and the LRS are reported in Fig.\ref{Fig:ResistanceRatios} for the different design options. As it can be seen, for the depletion-mode FeFET the high channel doping concentrations reduce the HRS over LRS ratio. Morever, the HRS over LRS ratio is smaller for the 80nm thick channel. This can be explained in terms of the effective thickness of the conductive channel that, for depletion-mode FETs, increases with t$_{ch}$. 

Fig.\ref{Fig:ResistanceRatios} also shows that for the enhancement-mode FeFETs the resistance modulation is less sensitive to t$_{ch}$, because in enhancement-mode transistors the current is carried by the electron inversion layer.

The large HRS/LRS ratios in Fig.\ref{Fig:ResistanceRatios} suggest that a large number of intermediate resistance levels may be accommodated in the devices, particularly in the enhancement-mode FeFETs. A systematic investigation of this latter aspect, however, would require to explore also the use of a fast, pulsed read operation \cite{Mulaosmanovic_EDL2020}, which goes beyond the quasi static regime considered so far, and actually beyond the scope of the present work.

\section{Conclusions}

We have investigated through numerical simulations and comparisons to experiments the ferroelectric switching in FeFETs and the resulting device operation. We found that the trapping in the ferroelectric-dielectric stack plays a crucial role for the polarization switching and thus for  the hysteresis of the \IDS -\VGS\ characteristics. Then we used such a physical insight to explore on a sound ground the design of BEOL compatible FeFETs with a thin polysilicon channel. While our present results have been limited to a quasi static operation, an extension to pulsed read conditions is an important extension that we foresee to undertake in the near future.

\vspace{3mm}
\noindent {\bf Acknowledgments}
This work was supported by the European Union through the BeFerroSynaptic project (GA:871737).

\end{document}